\documentclass[12pt,cite,twocolumn]{iopart}
\usepackage{graphicx,epsf}
\begin{document}

\title[Equivalence of quenched and annealed averaging in models of 
disordered polymers]{Equivalence of quenched and annealed averaging in models of disordered polymers}

\author{V. Blavatska}

\address{Institute for Condensed
Matter Physics of the National Academy of Sciences of Ukraine, 
79011 Lviv, Ukraine}
\ead{viktoria@icmp.lviv.ua}
\begin{abstract}
Equivalence of the influence of quenched and annealed disorder on scaling properties of long flexible polymer chains
is proved by analyzing the $O(m)$-symmetric field theory in polymer (de Gennes) limit $m\to 0$. Additional symmetry properties of the model in this limit are discussed.  
\end{abstract}

\pacs{61.25.hp, 11.10.Hi, 64.60.ae, 89.75.Da}
\submitto{Journal of Physics: Condensed Matter}
\maketitle

In real physical processes, one is often interested how structural obstacles (impurities) in the environment 
alter the behavior of a system. In polymer physics, of great importance is understanding of the behavior of macromolecules in the presence of structural disorder,
e.g., in colloidal solutions \cite{Pusey86}, in vicinity of microporous membranes \cite{Cannel80} or in a crowded environment
of biological cells \cite{Minton01}. Structural obstacles strongly effect the protein folding and aggregation \cite{Horwich,Winzor06,Kumar,Echeverria10}.

 Dealing with systems that display randomness of structure, one usually encounters two types of ensemble averaging,
treated as annealed and quenched disorder \cite{Brout59,Emery75}. 
In the former case, a characteristic time of impurities dynamics is comparable to relaxation times in the pure system and impurity variables 
are a part of the disordered system phase space, whereas in the latter case the impurities can be considered as
fixed and thus one needs to perform configurational average over an ensemble of disordered systems with different
realization of the disorder. In principle, the critical behavior of systems with quenched and
annealed disorder is quite different. For magnetic systems  with annealed disorder it was proved long ago \cite{Fisher68}, 
that when a heat capacity critical exponent of an undiluted system $\alpha_{pure}$ is positive, then any critical index $x_{annealed}$ 
of an annealed system is determined by those of the corresponding pure one $x_{pure}$ by simple relation (so-called Fisher renormalization):
\begin{equation}
x_{annealed}=\frac{x_{pure}}{1-\alpha_{pure}},\,\,\,\,\,\,\alpha_{annealed}=-\frac{\alpha_{pure}}{1-\alpha_{pure}}.  \label{fisher}
\end{equation}
The quenched disorder causes changes in the critical indexes also only if $\alpha_{pure}$ of corresponding system is positive 
(this statement is known as Harris criterion \cite{Harris74}); however, the quantitative influence is far not so trivial as in annealed case 
and serves as a subject of intensive studies (for a review, see e.g. \cite{Blavatska05}).

It is established, that some conformational statistical properties of long flexible polymer chains in good solvent
are governed by universal scaling exponents. As an example, the effective linear size such as averaged mean-squared end-to-end distance  of a 
long flexible polymer chain scales with the number of monomers (mass) of the chain $N$ as:
 \begin{equation}
\langle R^2 \rangle  \sim  N^{2\nu}, \label {RR}
\end{equation}
with a universal  exponent $\nu$ depending on space dimension $d$ only (e.g., the phenomenological Flory theory \cite{deGennes79} gives $\nu(d)=3/(d+2)$). Note that at $d=4$ the intrachain steric interactions are irrelevant, and the polymer behaves as an idealized Gaussian chain with $\nu_{Gauss}=1/2$.    
The relation of the polymer size exponent to the correlation length critical index of the $m$-component spin vector model 
in the formal limit $m\to 0$ was provided by  P.-G. de Gennes (a well-known de Gennes limit). This allows to apply the 
advanced field theory approach developed to study the critical behavior of magnetic systems,  to analyze the scaling properties of polymers.

Analysis of the influence of annealed disorder on the polymer size exponent encounters some controversies. As it was pointed out in Refs. 
\cite{Duplantier88,Thirumalai88,Bhattacharjee91}, an attempt to apply directly the Fisher renormalization (\ref{fisher}) fails in this case. Really, e.g. in $d=2$ for polymer system $\alpha_{pure}=2-d\nu(d=2)=1/2$ is positive, and according to (\ref{fisher}) one receives: $\nu_{annealed}=3/2$, which is unphysical: extension of a chain beyond its total length ($\nu>1$) is not possible. Note, however, that similar misunderstanding arises, when trying to apply the Harris criterion to analyze the influence of point-like uncorrelated quenched disorder on scaling properties of polymers: though both in $d=2$ and $d=3$ the corresponding values of $\alpha$ is positive, 
such a type of disorder does not alter the values of scaling exponents (it has been proved analytically by Kim \cite{Kim83} and supported in numerous numerical studies \cite{Kremer81,Woo91,Grassberger93,Barat95,Lee96}).    
Only correlated quenched disorder, leading to  appearance of pore-like structures of fractal nature (e.g. percolation clusters) was shown to 
influence the conformational properties of polymer macromolecules in a non-trivial manner  \cite{Ordemann00,Rintoul94,Blavatska08,Blavatska01}.  In a number of works \cite{Cherayil90,Wu91,Ippolito98,Patel03}
it is pointed out that the distinction between quenched and annealed averages for an
infinitely long single polymer chain is negligible. 
In the present study, we aim to confirm these suggestions by providing the simple arguments, based on refined field-theoretical approach.

We start with the Edwards continuous chain model \cite{desCloizeaux}, considering a linear polymer chain in a solution in the presence of structural obstacles   presented by a path $ r(s)$,
parameterized by $0\leq s\leq S$ ($S$ is also called the Gaussian surface).  
The partition function of the system is given by functional integral:
\begin{eqnarray}
&&{ Z}(S)=\int {\cal D} \{r\}\exp\left [-\frac{1}{2}\int_0^{S}\!\!{\rm d}s
\left(\frac{{\rm d} {\vec  r}(s)}{{\rm d} s}\right)^2-\right.\label{pure}\\
&&\left.-\frac{u_0}{2}
\int_0^{S}\!\!{\rm d}s'\int_0^{S}\!\!{\rm d}\,s{''}\,\delta(\vec{r}(s'')-\vec{r}(s{'}))- \int_0^{S}\!\!{\rm d}s\,V(\vec{r}(s))\right]\nonumber.
\end{eqnarray}
Here, an integration is performed over different  polymer path $ \vec{r}(s)$  configurations \cite{Kozitsky}. The first term in the exponent represents the chain connectivity, the
second term describes the short range excluded volume interaction with bare coupling constant $u_0$,
and the last term contains  random potential $V(\vec{r}(s))$  arising due to the presence of structural disorder.
Let us denote by ${\overline{(\ldots)}}$ the average over different realizations of disorder and introduce notation for the second moment: 
 \begin{equation} 
 {\overline{ V(\vec{r}(s))V(\vec{r}(s'))}} \equiv v_0 g(\vec{r}(s)-\vec{r}(s')),\label{avv0}
\end{equation}
with $v_0$ being some constant (the first moment of the distribution
 ${\overline { V({\vec {r}}(s))}}=0$).
The case of structural disorder in the form of point-like
uncorrelated defects corresponds to:
\begin{equation}
g_{uncor}(\vec{r}(s)-\vec{r}(s'))=\delta(\vec{r}(s)-\vec{r}(s'))\label{uncor}
\end{equation}
(Kronecker delta function), whereas another interesting situation when structural obstacles are
spatially correlated at large distances may correspond to \cite{Weinrib83}:
\begin{equation}
g_{cor}(\vec{r}(s)-\vec{r}(s'))=|\vec{r}(s)-\vec{r}(s')|^{-a},\,\,\,\,a>0.\label{cor}
\end{equation}

In order to average the free energy ${\cal F}$ of the system over
different configurations of {\it quenched} disorder,  the usual replica method  is used \cite{Emery75}, 
which is based on formal relation
\begin{equation}
{\overline {\cal F}}={\overline {\ln  Z}}=\lim\limits_{n\to 0}\frac{({\overline {Z^n}}-1)}{n}.
\end{equation}
For replicated partition sum of continious chain model ${\overline {Z^n(S)}}$ one receives:
\begin{equation}
{\overline {Z^n(S)}}=\prod_{i=1}^n \int {\cal D} \{r_i\}\,
 \rm{e}^{-H_{{\rm quen}}}
 \end{equation}
with an effective Hamiltonian:
\begin{eqnarray}
&&H_{{\rm quen}}= \frac{1}{2}\int_0^{S}\!\!{\rm d}s
\sum_{\alpha=1}^n\left(\frac{{\rm d} {\vec r_{\alpha}}(s)}{{\rm d} s}\right)^2+ \nonumber\\
&&+\frac{u_0}{2}\sum_{\alpha=1}^n
\int_0^{S}\!\!{\rm d}s'\int_0^{S}\!\!{\rm d}\,s{''}\,\delta(\vec{r}_{\alpha}(s'')-\vec{r}_{\alpha}(s{'}))-\nonumber\\
&&-\frac{v_0}{2}\sum_{\alpha,\beta=1}^n
\int_0^{S}\!\!{\rm d}s'\int_0^{S}\!\!{\rm d}\,s{''}\,g(\vec{r}_{\alpha}(s'')-\vec{r}_{\beta}(s{'})).\label{Hcon}
\end{eqnarray}
Here, Greek indices denote replicas and the last term, describing effective interaction between replicas, appears due to the presence of disorder in environment.

To take into account the {\it annealed} disorder, one deals with averaged partition sum in the form \begin{equation}{\overline {Z(S)}}=\int {\cal D} \{r\}\, \rm{e}^{-H_{{\rm ann}}}\end{equation} with an effective Hamiltonian:  
\begin{eqnarray}
&&H_{{\rm ann}}= \frac{1}{2}\int_0^{S}\!\!{\rm d}s
\left(\frac{{\rm d} {\vec r}(s)}{{\rm d} s}\right)^2+ \nonumber\\
&&+\frac{u_0}{2}
\int_0^{S}\!\!{\rm d}s'\int_0^{S}\!\!{\rm d}\,s{''}\,\delta(\vec{r}(s'')-\vec{r}(s{'}))- \nonumber\\
&&-\frac{v_0}{2}
\int_0^{S}\!\!{\rm d}s'\int_0^{S}\!\!{\rm d}\,s{''}\,g(\vec{r}(s'')-\vec{r}(s{'})).\label{Hconann}
\end{eqnarray}
We aim to show, that both model (\ref{Hcon}) and (\ref{Hconann}) are equivalent in renormalization group sense.

The model (\ref{Hcon}) may be mapped to a field theory
by a Laplace transform from the Gaussian surface
$S$ to the conjugated chemical potential variable (mass)
$\mu_0$ according to \cite{desCloizeaux,Schafer91}:
\begin{equation} \label{laplace}
 \hat{\cal Z}^n(\mu_0)=\int{\rm d}S
\exp[-\mu_0 S]{\cal Z}^n(S).
\end{equation}
Exploiting the analogy between the polymer problem and $O(m)$ symmetric field theory in the limit $m\to 0$
(de~Gennes limit) \cite{deGennes79}, it can be shown \cite{desCloizeaux} that the problem of disordered  polymer system is related to the $m=0$-component field theory with an effective Lagrangean:
\begin{eqnarray}
 &&{\cal L}= \frac{1}{2} \sum_{\alpha=1}^{n}
\int{\rm d}^d x \left[\left(\mu_0^2|\vec{\phi}_{\alpha}(x)|^2+
|\nabla\vec{\phi}_{\alpha}(x)|^2\right)\right.+\nonumber\\
&&\left.+\frac{u_0}{4!}\left({\vec{\phi}_{\alpha}}^2(x)\right)^2\right]-\nonumber\\
&&-\frac{v_0}{4!}\sum_{\alpha,\beta=1}^{n}
\int{\rm d}^dx\,{\rm d}^dy\, g(x-y)
{\vec{\phi}_{\alpha}}^2(x){\vec{\phi}_{\beta}}^2(y).
\label{Leff}
\end{eqnarray}
Here, each $\vec{\phi}_{\alpha}$ is an $m$-component vector field
$\vec{\phi}_{\alpha}=(\phi_1^{\alpha},\ldots,\phi_m^{\alpha})$ and the last term describing replicas coupling contains the correlation
function in the form (\ref{uncor}) or (\ref{cor}).

Usually, the effective Lagrangean (\ref{Leff})  is studied in the limiting case $n \to 0$ 
that corresponds to quenched disorder. Let us note, however, that another 
limit $ n \to \infty $ has physical interpretation, too. Indeed, as it was shown 
in \cite{Emery75}, this limit corresponds to the annealed disorder. Therefore, the 
problem of analysis of scaling properties of a polymer chain in two different 
types of disorder is reduced to study of the critical properties of the field 
theory (13) in the polymer limit $ m=0$ and two different limits $n \to 0$ (quenched 
disorder) and $n \to \infty$ (annealed disorder). 


One of the ways of extracting the scaling behavior of the model~(\ref{Leff})
is to apply the field-theoretical renormalization
group~(RG) method \cite{rgbooks} in the massive scheme, with the  Green's functions renormalization at
non-zero mass and zero external momenta. The bare Green's function
 $G_0^{(N)}$ can be defined as an average of $N$ field components performed with the corresponding effective Lagrangean ${\cal L}$:
\begin{eqnarray}
&&\delta\left(\sum   \vec{p}_j\right)G_0^{(N)}(\{\vec{p}\}; \mu^2_0;\{\lambda_0\}) =\nonumber\\
&&\int^{\Lambda_0}\!\!{\rm e}^{i {p}_j {r}_j}\langle\vec{\phi}(r_1)\dots{\vec{\phi}}(r_N)
\rangle^{{\cal  L}}
{\rm d}^d r_1 \dots {\rm d}^d
r_N\,,
\end{eqnarray}
here,  $\{\vec{p} \}=(\vec{p}_1\,,\ldots,\vec{p}_N)$ is the sets of
external momenta, $\{\lambda_0\}=(u_0,v_0)$ is the set of bare coupling constants  and $\Lambda_0$ is a cut-off.
\begin{figure}[!t]
\begin{center}
\includegraphics[width=8cm]{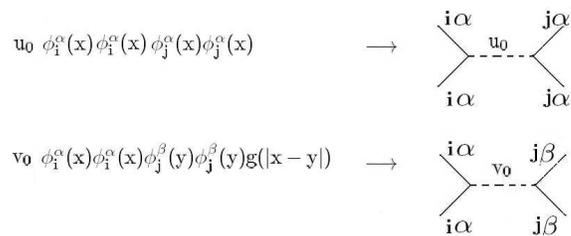}
\end{center}
\caption{\label{fig1} Diagrammatic representation of quartic potentials in (\ref{Leff}).}
\end{figure}
Let us start by  representing the two quartic potentials in (\ref{Leff}) in the form of four-leg vertices accordingly to so-called ``faithful representation" \cite{rgbooks}, as shown in Fig. 1. This allows convenient representation of the terms in perturbation theory expansions of   
the bare Green functions in coupling $u_0$ and $v_0$ by diagrams, given in Fig. 2. The arguments given below are valid in any order of 
the perturbation theory. However, for simplicity, we restrict ourselves by the second-order expansion of a function $G_0^2({{p}})$  (so-called ``two-loop" approximation: in this case, only those diagrammatic contributions are taken into account, which contain no more than two closed loops):   
 \begin{eqnarray}
  &&G_0^2({{p}})=1-\frac{u_0}{6}\left( 2I_{u_0}^{(b)}+mI_{u_0}^{(c)}\right)+\frac{v_0}{6}\left( 2I_{v_0}^b+mnI_{v_0}^c\right)\nonumber\\
&& +\frac{u_0^2}{9} \left( 2m^2I_{u_0^2}^{(d)}+4mI_{u_0^2}^{(e)}+ 4mI_{u_0^2}^{(f)}+8I_{u_0^2}^{(g)}+8I_{u_0^2}^{(i)}+4mI_{u_0^2}^{(h)}\right)\nonumber\\
&&-\frac{2u_0v_0}{9}\left(2m^2nI_{u_0v_0}^{(d)}+2mnI_{u_0v_0}^{(e)}+2mI_{v_0u_0}^{(e)}+ 2mnI_{u_0v_0}^{(f)}\right.\nonumber\\
&&+ 2mI_{v_0u_0}^{(f)}+8I_{u_0v_0}^{(g)}\nonumber\\
&&\left.+8I_{u_0v_0}^{(i)}+4mI_{u_0v_0}^{(h)} \right)+\frac{v_0^2}{9}\left( 2m^2n^2I_{v_0^2}^{(d)}+4mnI_{v_0^2}^{(e)}\right.\nonumber\\
&&{ +}\left. 4mnI_{v_0^2}^{(f)}+8I_{v_0^2}^{(g)}+8I_{v_0^2}^{(i)}+4mnI_{v_0^2}^{(h)} \right).\label{ggg}
 \end{eqnarray}
 Here, the following notation is used: $I_{u_0^kv_0^l}^{(z)}$ denotes loop integral, corresponding to diagram (z) in Fig. 2, containing $k$ vertices 
 of the type $u_0$
 and $l$ vertices of the type $v_0$. The loop integrals are dependent on the type of quartic interaction, e.g. in the case of point-like uncorrelated disorder  (\ref{uncor}), the integrals corresponding to diagram (e) read: 
 \begin{eqnarray}
 I_{u_0^2}^{(e)}=I_{v_0^2}^{(e)}=I_{u_0v_0}^{(e)}=\frac{1}{(p^2+\mu_0^2)^2}\!\int\!\!\frac{{\rm d}\, {\vec{q}_1}\,\,{\rm d}\, {\vec{q}_2}}{(q_1^2+\mu_0^2)^2(q_2^2+\mu_0^2)},\nonumber
 \end{eqnarray} 
whereas in the case of long-range correlated disorder, corresponding to (\ref{cor}), one has \cite{Blavatska01}:
\begin{eqnarray}
&& I_{u_0^2}^{(e)}=\frac{1}{(p^2+\mu_0^2)^2}\!\int\!\!\frac{{\rm d}\, {\vec{q}_1}\,\,{\rm d}\, {\vec{q}_2}}{(q_1^2+\mu_0^2)^2(q_2^2+\mu_0^2)},\nonumber\\
&& I_{v_0^2}^{(e)}=\frac{1}{(p^2+\mu_0^2)^2}\!\int\!\!\frac{{\rm d}\, {\vec{q}_1} |\vec{q}_1|^{a-d}\,\,{\rm d}\, {\vec{q}_2}|\vec{q}_2|^{a-d}}{(q_1^2+\mu_0^2)^2(q_2^2+\mu_0^2)},\nonumber\\
&& I_{u_0v_0}^{(e)}=I_{v_0u_0}^{(e)}=\frac{1}{(p^2+\mu_0^2)^2}\!\int\!\!\frac{{\rm d}\, {\vec{q}_1} |\vec{q}_1|^{a-d}\,\,{\rm d}\, {\vec{q}_2}}{(q_1^2+\mu_0^2)^2(q_2^2+\mu_0^2)}.\nonumber \end{eqnarray}
 \begin{figure}[!b]
\begin{center}
\includegraphics[width=8cm]{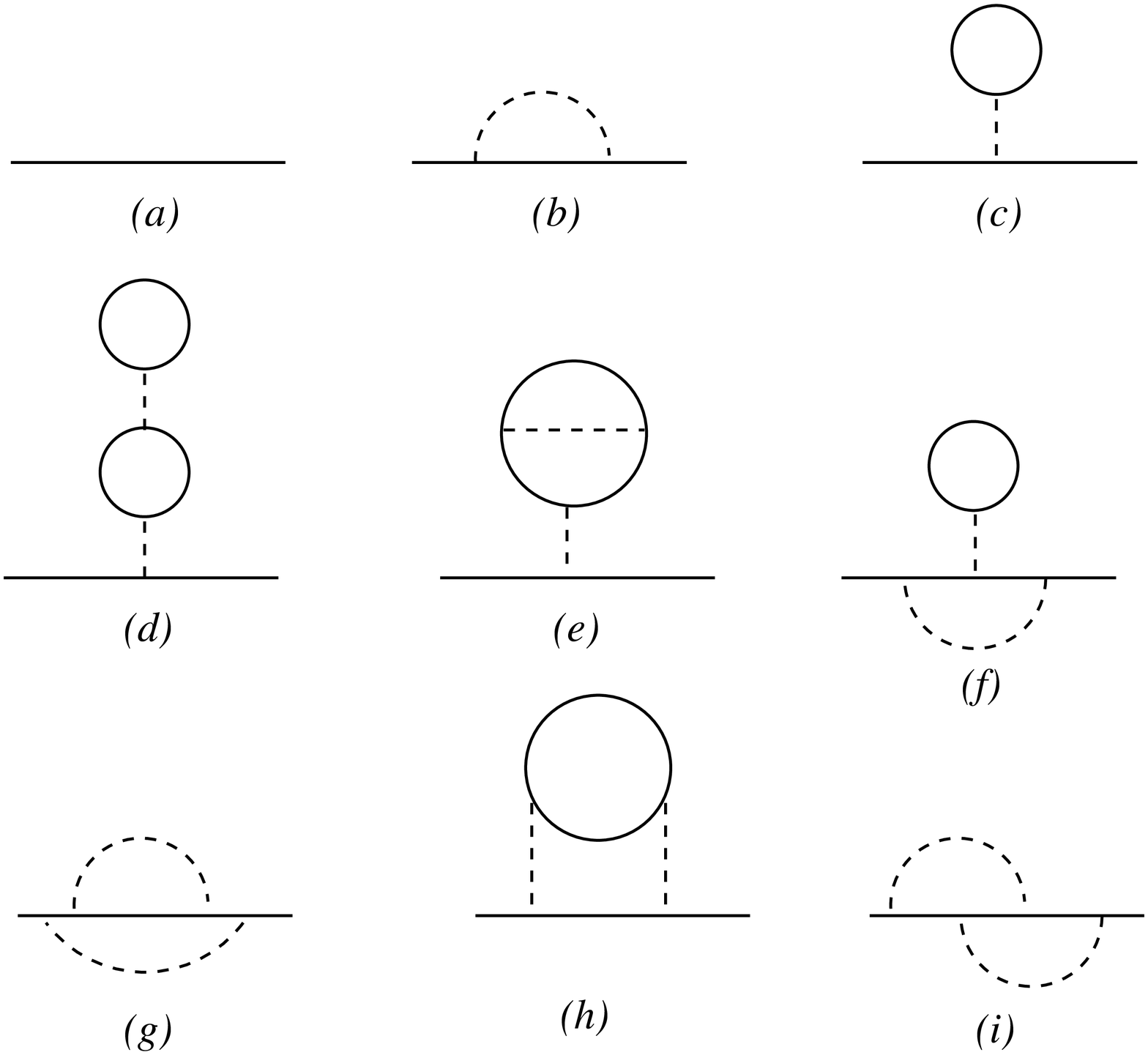}
\end{center}
\caption{\label{fig2} Contributions into the Green's function $G_0^{(2)}$ up to two-loop approximation. Solid lines denote propagators $\mu_0^2+q^2$,  loops imply integration over 
internal momenta $\vec{q}$. Every interactive diagram appears twice, once with each of the two couplings $u_0$ and $v_0$, respectively.}
\end{figure}
 The combinatorial prefactors corresponding to each loop integral in series (\ref{ggg}) are defined, however, only by the topological type of corresponding diagram. One may easily check, that the diagrams, containing closed loops, produce combinatorial prefactors, 
proportional to $m$ or $mn$ \cite{Kim83}. It is crucial to note, that $n$ appears {\it only} in combinations, where it is multiplied by $m$. The same feature 
is observed also in analyzing any other Green function $G^N$ in any order of perturbation theory in coupling constants $u_0$, $v_0$. 
Thus, when the polymer limit $m\to 0$ is implied, all the renormalization group functions simultaneously become $n$-independent.  
Recalling  discussion below (\ref{Leff}) concerning dependence of the type of averaging on the choice of 
replica parameter $n$, this leads us to straightforward conclusion: as long as the polymer problem is under interest, both quenched ($n\to 0$) and
annealed ($n\to\infty$) averaging are equivalent.  

Let us now recall once more the effective Hamiltonian, describing continuous polymer chain in disordered environment (\ref{Hcon}) and corresponding field-theoretical model (\ref{Leff}).  The conclusion about $n$-independence in polymer limit alows us to put $n=1$ directly into these expressions, obtaining respectively:
\begin{eqnarray}
&&H_{{\rm quen}}= \frac{1}{2}\int_0^{S}\!\!{\rm d}s
\left(\frac{{\rm d} {\vec r}(s)}{{\rm d} s}\right)^2+ \nonumber\\
&&+\frac{u_0}{2}\int_0^{S}\!\!{\rm d}s'\int_0^{S}\!\!{\rm d}\,s{''}\,\delta(\vec{r}(s'')-\vec{r}(s{'}))-\nonumber\\
&&-\frac{v_0}{2}
\int_0^{S}\!\!{\rm d}s'\int_0^{S}\!\!{\rm d}\,s{''}\,g(\vec{r}(s'')-\vec{r}(s{'}));\label{Hcon1}\\
 &&{\cal L}= \frac{1}{2} 
\int{\rm d}^d x \left[\left(\mu_0^2|\vec{\phi}(x)|^2+
|\nabla\vec{\phi}(x)|^2\right)\right.+\nonumber\\
&&\left.+\frac{u_0}{4!}\left({\vec{\phi}}^2(x)\right)^2\right]-\nonumber\\
&&-
\frac{v_0}{4!}\int{\rm d}^dx\,{\rm d}^dy\, g(x-y)
{\vec{\phi}}^2(x){\vec{\phi}}^2(y).\label{Leff1}
\end{eqnarray}
One immediately reveals the equivalence of (\ref{Hcon1}) with corresponding Hamiltonian of the system with annealed disorder (\ref{Hconann}). 
In performing analytical calculations one can thus restrict oneself to the simpler case of annealed averaging, which considerably simplifies the 
renormalization group procedure. In particular, in the case of point-like uncorrelated defects (taking correlation function in the form (\ref{uncor}))
one then  notices the equivalence of the last two terms in both (\ref{Hcon1})  and (\ref{Leff1}),  which  in turn allows to adsorb the third term (arising 
due the presence of disorder) into the second term (describing the excluded volume effect) by simple redefinition of coupling constant $u_0\equiv u_0-v_0$.   Thus,  in the renormalization group sence the presence of uncorrelated 
point-like defects at low densities does not influence the scaling properties of polymers. This conclusion was obtained earlier for the case of polymers in quenched disorder by Kim  \cite{Kim83} on the basis of a much more refined field-theoretical study. 
      
 \section*{Acknowledgment}
This work was supported in part by the FP7 EU IRSES project N269139
``Dynamics and Cooperative Phenomena in complex Physical and
Biological Media''.


\section*{References}
\begin{thebibliography}{10}

\bibitem{Pusey86}
 Pusey P N  and  van Megen W  1986 Nature {\bf 320}  340

\bibitem{Cannel80}
  Cannell D S and  Rondelez F  Macromolecules 1980 {\bf 13}   1599      

\bibitem{Minton01}
  Record M T,   Courtenay E S,     Cayley S, and   Guttman H J 1998  Trends. 
Biochem.  Sci.  {\bf 23}   190;
 Minton  A P 2001  J.  Biol.  Chem.   {\bf 276} 10577;  Ellis R J  and   Minton A P 2003 Nature  {\bf 425}   27      

\bibitem{Horwich}
 Horwich  A 2004 Nature {\bf 431}   520      

\bibitem{Winzor06}
 Winzor D J  and   Wills P R 2006 Biophys.  Chem.  {\bf 119}   186;
  Zhou H -X,   Rivas G,   and   Minton A P 2008 Annu.  Rev.  Biophys.  {\bf 37}   375      

\bibitem{Kumar}
  Kumar S,    Jensen  I,   Jacobsen J L, and   Guttmann A J 2007   Phys.  Rev.  Lett.  {\bf 98}   128101;
 Singh A R,    Giri D  and   Kumar S 2009  Phys.  Rev.  E {\bf 79}   051801     


\bibitem{Echeverria10}
 Echeverria C  and   Kaprai R  2010 J.  Chem.  Phys.  {\bf 132}   104902     

\bibitem{Brout59}
 Brout R  1959  Phys.  Rev.  {\bf 115}      824 

\bibitem{Emery75}
  Emery V  J  1975  Phys.  Rev.  B {\bf 11}    239;   Edwards S  F and  Anderson  P W 1975   J.  Phys.  F {\bf 5}   965    

\bibitem{Fisher68}
  Fisher M  E  1968 Phys.  Rev.  {\bf 176}   257     


\bibitem{Harris74}
  Harris A B  1974  J.  Phys.  C {\bf 7}   1671    

%

\bibitem{Blavatska05}
 Folk R, Holovatch Yu,	 and   Yavorski T  2003  Physics-Uspiekhi {\bf 46} 169;  Holovatch Yu,	  Blavats'ka V, Dudka M,   von Ferber C,      Folk R, and   Yavorski T  2002 J. Mod. Phys. B {\bf 16}  4027



\bibitem{deGennes79}
de Gennes P G    1979 {\it Scaling Concepts in Polymer Physics } (Ithaca: Cornell
University Press)        

\bibitem{Duplantier88}
 Duplantier B   1988 Phys.  Rev.  A   {\bf 38}     3647; 
B  Duplantier and H  Saleur   1987 Phys.  Rev.  Lett.  {\bf 59}     539

\bibitem{Thirumalai88}
 Thirumalai D  1988 Phys.  Rev.  A {\bf 37}          269 

\bibitem{Bhattacharjee91}
  Bhattacharjee S  M and Chakrabarti  B  K  1991  Europhys.  Lett.  {\bf 15}   259    

\bibitem{Kim83}
 Kim Y  1983  J.  Phys.  C {\bf 16}   1345     


\bibitem{Kremer81}
 Kremer K 1981    Z.   Phys.   B {\bf 45}  149 

\bibitem{Woo91}
 ~Lee S ~B and ~Nakanishi H  1988   Phys.   Rev.   Lett.   {\bf 61}  2022; ~Lee  S ~B,   ~Nakanishi  H,   and  Kim Y  1989     Phys.   Rev.   B {\bf 39}  9561;
 ~Woo K ~Y and  ~Lee S ~B  1991   Phys.   Rev.   A {\bf 44}  999  

\bibitem{Grassberger93}
~Grassberger P   1993   J.   Phys.   A {\bf 26}  1023 

\bibitem{Barat95}
 Barat K and   Chakrabarti B  K 1995  Phys.  Reports   {\bf 258}   377     


\bibitem{Lee96} 
 ~Lee S ~B 1996    J   Korean  Phys   Soc   {\bf 29}  1


\bibitem{Ordemann00}
  Ordemann A,   Porto  M,  and   Roman H E 2000 Phys.  Rev.  E {\bf 65}   021107;
 2002  J.  Phys.  A {\bf 35}   8029     

\bibitem{Rintoul94}
  ~Rintoul M D,  ~Moon J   and  ~Nakanishi H  1994 Phys.  Rev.  E {\bf 49}  
2790     

\bibitem{Blavatska08}
 Blavatska V and  Janke  2008 W  Europhys.  Lett.  {\bf 82}   66006;
2008  Phys.  Rev.  Lett.  {\bf 101}    125701;
2009 J.  Phys.  A {\bf 42}   015001      



\bibitem{Blavatska01}
Blavats'ka~ V,     von Ferber~C, and     Holovatch~Yu   2001   J. ~Mol.  Liq      {\bf 91}   77; 2001 Phys.  Rev. ~E     
{\bf 64}   041102; 2010 Phys.  Lett.  A      {\bf 374}   2861 


\bibitem{Cherayil90}
  Cherayil B J 1990 J.  Chem.  Phys.  {\bf 92}   6246      

\bibitem{Wu91}
  Wu D,    Hui K  and  Chandler D 1991   J.  Chem.  Phys.  {\bf 96}   835     

\bibitem{Ippolito98}
 Ippolito I,   Bideau  D  and  Hansen A  1998 Phys.  Rev.  E {\bf 57}   3656      

\bibitem{Patel03}
 Patel D M  and   Fredrickson G H, 2003 Phys.  Rev.  E {\bf 68}   051802      




\bibitem{desCloizeaux}
 desCloizeaux J and Jannink  G  1990
{\it Polymers in Solution: Their Modeling and Structure}   (Oxford:  Clarendon Press)     

\bibitem{Kozitsky} See e. g.  Albeverio~S,      Kondratiev~Yu,      Kozitsky~Yu,  and    R\"ockner~M   
2009 {\it The Statistical Mechanics of Quantum Lattice Systems  A~Path
Integral Approach}  (Z\"urich: European Math.  Soc.  Publishing House)  
  
\bibitem{Weinrib83}
 Weinrib A and    Halperin  B I 1983 
Phys.  Rev.  B {  27}     413 

  


\bibitem{Schafer91}
Sch\"afer L and     Kapeller C   1985  J. ~Phys.       {\bf 46}    1853; 
1990 Colloid  Polym.  Sci.       {\bf 268}    995. 







 \bibitem{rgbooks}
Amit~D J   1989  {\it Field T	heory   the Renormalization Group   and Critical
Phenomena}  ( Singapore: World Scientific);
Zinn-Justin~J   1996 {\it Quantum Field Theory and Critical Phenomena} 
( Oxford: Oxford University Press);
Kleinert H and     Schulte-Frohlinde~V   2001  {\it Critical Properties of
$\phi^4$-Theories} (Singapore:   World Scientific)      
\end{thebibliography}
\end{document}